\input{aipcheck}

\documentclass[
    ,final            
  ]
  {aipproc}

\layoutstyle{6x9}

\begin{document}

\title{DMTPC: A dark matter detector with directional sensitivity}


\author{J.B.R.~Battat}{
  address={Massachusetts Institute of Technology, Cambridge, MA 02139, USA}
}
\author{S. Ahlen}{
  address={Boston University, Boston, MA, 02215, USA}
}
\author{T.~Caldwell}{
  address={Massachusetts Institute of Technology, Cambridge, MA 02139, USA}
}
\author{D.~Dujmic}{
  address={Massachusetts Institute of Technology, Cambridge, MA 02139, USA}
}
\author{A.~Dushkin}{
  address={Brandeis University, Waltham, MA, 02453, USA}
}
\author{P.~Fisher}{
  address={Massachusetts Institute of Technology, Cambridge, MA 02139, USA}
}
\author{F.~Golub}{
  address={Brandeis University, Waltham, MA, 02453, USA}
}
\author{S.~Goyal}{
  address={Brandeis University, Waltham, MA, 02453, USA}
}
\author{S.~Henderson}{
  address={Massachusetts Institute of Technology, Cambridge, MA 02139, USA}
}
\author{A.~Inglis}{
  address={Boston University, Boston, MA, 02215, USA}
}
\author{R.~Lanza}{
  address={Massachusetts Institute of Technology, Cambridge, MA 02139, USA}
}
\author{J.~Lopez}{
  address={Massachusetts Institute of Technology, Cambridge, MA 02139, USA}
}
\author{A.~Kaboth}{
  address={Massachusetts Institute of Technology, Cambridge, MA 02139, USA}
}
\author{G.~Kohse}{
  address={Massachusetts Institute of Technology, Cambridge, MA 02139, USA}
}
\author{J.~Monroe}{
  address={Massachusetts Institute of Technology, Cambridge, MA 02139, USA}
}
\author{G.~Sciolla}{
  address={Massachusetts Institute of Technology, Cambridge, MA 02139, USA}
}
\author{B.N.~Skvorodnev}{
  address={Brandeis University, Waltham, MA, 02453, USA}
}
\author{H.~Tomita}{
  address={Boston University, Boston, MA, 02215, USA}
}
\author{R.~Vanderspek}{
  address={Massachusetts Institute of Technology, Cambridge, MA 02139, USA}
}
\author{H.~Wellenstein}{
  address={Brandeis University, Waltham, MA, 02453, USA}
}
\author{R.Yamamoto}{
  address={Massachusetts Institute of Technology, Cambridge, MA 02139, USA}
}



\begin{abstract}
By correlating nuclear recoil directions with the Earth's direction of
motion through the Galaxy, a directional dark matter detector can
unambiguously detect Weakly Interacting Massive Particles (WIMPs),
even in the presence of backgrounds.  Here, we describe the Dark
Matter Time-Projection Chamber (DMTPC) detector, a TPC filled with
CF$_4$ gas at low pressure (0.1~atm).  Using this detector, we have
measured the vector direction (head-tail) of nuclear recoils down to
energies of 100~keV with an angular resolution of $\leq$15$^\circ$.
To study our detector backgrounds, we have operated in a basement
laboratory on the MIT campus for several months.  We are currently
building a new, high-radiopurity detector for deployment underground
at the Waste Isolation Pilot Plant facility in New Mexico.
\end{abstract}

\maketitle

\section{Directional dark matter detection}
Astrophysical observations, coupled with simulations of galaxy
formation, tell us that the baryonic disk of the Milky Way is likely
embedded in a much larger halo of dark matter.  The stars and gas are
believed to rotate with respect to the halo.  An Earth-bound observer,
in orbit about the center of the Galaxy, will therefore move through
the dark matter distribution and experience a head-wind of dark matter
particles.

The smoking-gun signature of the above paradigm is an O(1) daily
modulation\footnote{Here, daily means sidereal day (with respect to
distant stars), not solar day (with respect to the Sun).} in the
direction of arrival of the dark matter wind at the Earth, due to the
Earth's rotation \cite{spergelPRD1988}.  For a detector at latitude
$\sim$45$^\circ$ (e.g. at MIT in Cambridge, MA), if the wind appears
to come from the horizon at time $t$, then at $t+12$~h, the wind will
come from overhead.

The goal, then, of a directional detector is to reconstruct the
time-dependent direction of the dark matter wind by measuring the
recoil axis and direction of the recoiling nuclei.  A large asymmetry
in the direction of nuclear recoils with respect to the Galaxy would
be the smoking gun for dark matter detection since no known background
source can mimic this modulation signal.  Here, we describe the Dark
Matter Time-Projection Chamber (DMTPC) apparatus, a gas-based detector
that is sensitive to the direction of the dark matter wind.

\section{Detector description and performance}
\begin{figure}
  \includegraphics[height=.4\textheight]{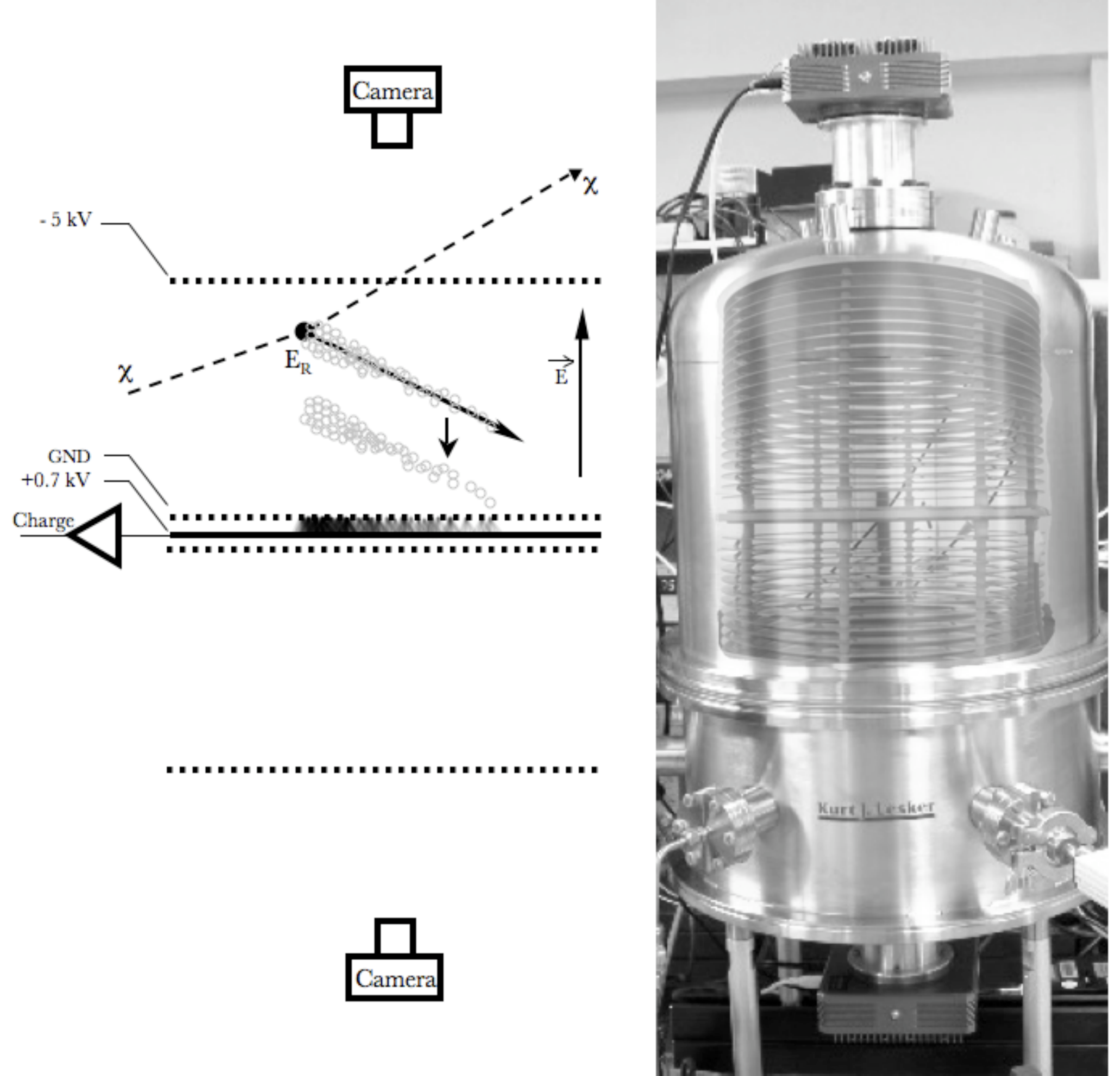}
  \hspace{0.5in}
  \includegraphics[height=.6\textwidth]{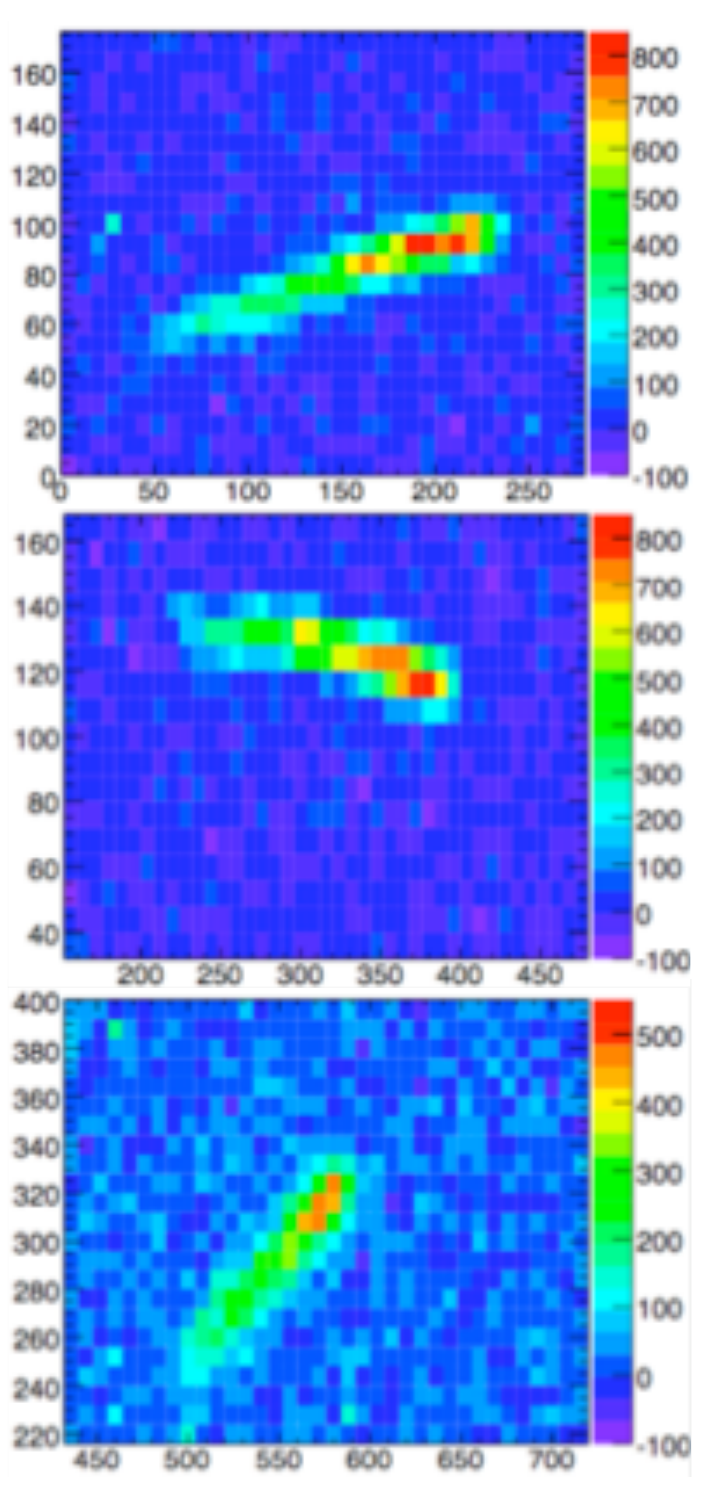}
  \caption{\label{fig:detector}({\em left}) Schematic of a WIMP-nucleus interaction in one side of the dual TPC.  (center) Photograph of the DMTPC 10L detector with an image of the field cage overlaid to provide an artificial glimpse inside the vacuum vessel.  ({\em right}) CCD images of $^{252}$Cf neutron-induced nuclear recoils \cite{dujmicAstroPart2008}.}
\end{figure}

The DMTPC detector, shown in Figure \ref{fig:detector}, is a dual,
back-to-back, time-projection chamber (TPC) filled with CF$_4$ gas at
low pressure (0.1~atm), with electronic and optical (CCD and PMT)
readout.  The active volume of the detector is 10~L, which, at
0.1~atm, corresponds to 3.3~g of CF$_4$.

A recoiling nucleus from a dark matter interaction will ionize the gas
along its track.  The liberated electrons drift under a uniform
electric field toward an amplification region.  The 20~cm long drift
region is established by the cathode at $\sim-5$~kV and the ground
plane.  These conductive planes are fine-woven meshes made from
28~$\mu$m wire with a 256~$\mu$m pitch.  The amplification region
consists of a copper-clad G10 anode at +0.7~kV separated from the
ground plane by 500~$\mu$m.  We achieve typical gas gains of $10^5$
with minimal sparking.  The mesh-based amplification region produces
two-dimensional images of particle tracks.

The charge deposited on the anode, which is proportional to the total
recoil energy, is recorded with a fast digitizer.  The scintillation
light from tracks is imaged with a CCD camera, providing a measurement
of the total recoil energy, the energy loss per unit length $dE/dx$,
and the length and shape of the 2-dimensional projection of the track.
A photomultiplier tube (PMT) also detects the scintillation light.
The timing profile of the PMT signal provides information about the
third dimension (vertical extent) of the track.  The combination of
electronic and light readout ensures effective background
discrimination.


The energy resolution of the charge readout is $\sim 10$\% at 5.9~keV
(measured with an $^{55}$Fe source), and is $\sim 15$\% at 50~keV for
the CCD readout.  The detector has excellent electron rejection
(better than $10^6$) owing to the low surface brightness and extensive
length of electron tracks.  Alpha tracks are distinguished from
nuclear recoils by their energy-range relationship.

\section{CF$_4$ gas properties}
CF$_4$ is an excellent target gas for a dark matter detector.  It
provides strong sensitivity to spin-dependent interactions because of
the unpaired proton in fluorine.  It is also an efficient
scintillator, with significant emission around
650~nm \cite{kabothCF4}, which is well matched to the quantum
efficiency of the ubiquitous silicon CCDs.

CF$_4$ also has very low electron diffusion.  In order to faithfully
reconstruct a recoil track, the transverse electron diffusion must not
exceed the recoil track length.  Typical WIMP-induced recoils in our
detector extend 1--3 millimeters.  Recent measurements by our group
have shown that we can drift electrons over 20~cm with less than 1~mm
of transverse diffusion \cite{caldwell2009}.

\section{Head-tail measurements for nuclear recoils}
The angular distribution of WIMP-induced fluorine nuclei recoils is
similar to those induced by $^{252}$Cf neutrons.  With 75~torr of
CF$_4$ in the TPC, we used a $^{252}$Cf neutron source to calibrate the
sensitivity of our detector and analysis software to WIMP
events \cite{dujmicAstroPart2008}.  Sample images of nuclear recoil
candidate events are shown in Figure \ref{fig:detector}.  The recoil
axis is manifest in the track topology, and the vector direction of
the recoil (the head-tail effect) is clearly determined from the light
distribution along the track: higher light intensity marks the start
of the recoil.  It is readily apparent from the ensemble of images
that the $^{252}$Cf neutrons were incident from the right.

\begin{figure}
  \includegraphics[height=.3\textwidth]{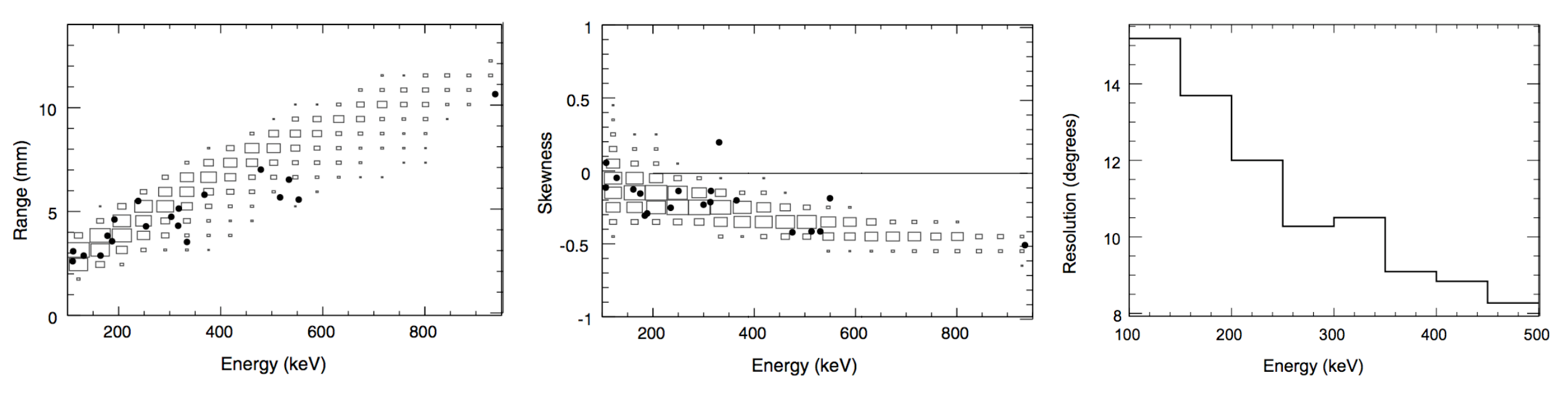}
  \caption{\label{fig:rangeSkewness} ({\em left}) The range vs. energy for candidate nuclear recoils (points) and the Monte Carlo prediction (box histogram).  ({\em center}) The skewness parameter $S$ vs. energy for data and Monte Carlo.  We can correctly determine the head-tail down to 100~keV.  ({\em right}) Monte Carlo study of the angular resolution of the track reconstruction.  Images taken from \cite{dujmicAstroPart2008}.}
\end{figure}

A plot of the observed range vs. energy for candidate nuclear recoil
events is shown in Figure \ref{fig:rangeSkewness}.  The observations
agree very closely with our Monte Carlo studies.  We quantify our
ability to reconstruct the head-tail of nuclear recoils with a
``skewness'' parameter $S=\mu_3/\mu_2^{3/2}$, where $\mu_2$ and
$\mu_3$ are the second and third moments, respectively, of the light
distribution along the track.  For the neutron calibration run,
kinematic constraints require that $S$ be negative.
Figure \ref{fig:rangeSkewness} shows that we successsfully reconstruct
the head-tail for nuclear recoils down to 100~keV.  Furthermore, Monte
Carlo studies show that we reconstruct the nuclear recoil direction
with an angular resolution of 15$^\circ$ at 100~keV, improving to
10$^\circ$ at 300~keV (Figure \ref{fig:rangeSkewness}).

\section{Current and future plans}
The 10L DMTPC detector was run in a basement laboratory on the MIT
campus for nine weeks to study the detector backgrounds.  The chamber
was refilled with CF$_4$ gas each day to ensure a gain stability of
1\%.  An analysis of the data from this surface run will be the
subject of a forthcoming publication.

We are currently constructing a second detector to deploy underground
at the Waste Isolation Pilot Plant (WIPP).  At WIPP (1.6 km.w.e.), we
expect much less than one neutron-induced background per year in our
detector.  In addition, careful attention to material radiopurity and
environmental radon levels during assembly should strongly suppress
the alpha backgrounds.

Meanwhile, we are also designing a cubic meter detector comprised of
four TPC volumes and transparent mesh anodes.  At 75~torr, the cubic
meter detector would contain 0.38~kg of target material.  With three
months of live time (exposure $\sim$0.1~kg~yr), this detector is
capable of setting leading constraints on spin-dependent WIMP-proton
interactions \cite{dujmicAstroPart2008}.

\begin{theacknowledgments}
This work is supported by the Advanced Detector Research Program of the U.S. Department of Energy, the National Science Foundation, the Reed Award Program, the Ferry Fund, the Pappalardo Fellowship program, the MIT Kavli Institute for Astrophysics and Space Research, and the MIT Physics Department.
\end{theacknowledgments}



\bibliographystyle{aipproc}   

\bibliography{dmrefs}

\end{document}